\documentclass[aps,prd,twocolumn,prd,superscriptaddress,nofootinbib]{revtex4-1}

\usepackage[utf8]{inputenc}

\usepackage{mathtools}
\usepackage{amsfonts}
\usepackage{mathrsfs}
\usepackage{bbm}
\usepackage{slashed}

\usepackage{hyperref}
\hypersetup{
    pdftitle={},
    colorlinks=true,     
    linkcolor=blue,      
    citecolor=blue,      
    filecolor=blue,      
    urlcolor=blue        
}
\usepackage{cleveref}
\Crefname{equation}{Eq.}{Eqs.}
\Crefname{figure}{Fig.}{Figs.}
\Crefname{tabular}{Tab.}{Tabs.}

\usepackage{xcolor}
\usepackage{enumitem}
\hypersetup{colorlinks   = true, 
	urlcolor     = black, 
	linkcolor    = blue, 
	citecolor   = teal}

\usepackage{units}

\usepackage[utf8]{inputenc}

\usepackage{soul}

\renewcommand{\Re}{\textrm{Re}}

\newcommand{\SU}{{\mathrm{SU}}}

\newcommand{\pp}{{\mathbf{p}}}
\newcommand{\pb}{{\mathbf{\bar{p}}}}
\newcommand{\zerob}{{\mathbf{\bar{0}}}}
\newcommand{\ptilde}{{\mathbf{\tilde{p}}}}
\newcommand{\xx}{{\mathbf{x}}}
\newcommand{\xb}{{\mathbf{\bar{x}}}}
\newcommand{\yb}{{\mathbf{\bar{y}}}}
\newcommand{\yy}{{\mathbf{y}}}
\newcommand{\EE}{{\mathbf{E}}}

\newcommand{\Tr}{{\mathrm{Tr}}}
\newcommand{\dd}{{\mathrm{d}}}
\newcommand{\su}{\mathfrak{su}}

\newcommand{\jhat}{\hat{\mathbf{j}}}
\newcommand{\khat}{\hat{\mathbf{k}}}
\newcommand{\rr}{\mathbb{R}}
\newcommand{\cc}{\mathbb{C}}
\newcommand{\Dscr}{\mathscr{D}}
\newcommand{\Pscr}{\mathscr{P}}
\newcommand{\Ccal}{\mathcal{C}}
\newcommand{\Gcal}{\mathcal{G}}
\newcommand{\Ocal}{\mathcal{O}}

\newcommand{\tot}{\mathrm{tot}}


\begin{document}

\title{Condensation and prescaling in spatial Polyakov loop correlations far from equilibrium}

\author{Daniel Spitz}
\email{daniel.spitz@mis.mpg.de}
\affiliation{Max Planck Institute for Mathematics in the Sciences, Inselstraße 22, Leipzig, 04103, Germany}
\affiliation{Institute for Theoretical Physics, Heidelberg University, Philosophenweg 16, 69120 Heidelberg, Germany}

\author{Kirill Boguslavski}
\affiliation{Institute for Theoretical
  Physics, Technische Universit\"{a}t Wien, 1040 Vienna, Austria}

\author{Thimo Preis}
\affiliation{Institute for Theoretical Physics, Heidelberg University, Philosophenweg 16, 69120 Heidelberg, Germany}

\begin{abstract}
    The far-from-equilibrium dynamics of spatial Polyakov loop correlations, which provide gauge-invariant observables akin to effective particle numbers for gluon plasmas, are investigated within real-time $\SU(N_c)$ lattice gauge theory at weak couplings and large gluon occupations.
    The momentum zero mode of these correlations reveals the dynamic formation of a condensate, while at nonzero momenta, energy is transported toward the ultraviolet.
    We demonstrate that the non-zero momentum dynamics is well described by a direct cascade in terms of gauge-invariant Polyakov loop excitations, exhibiting self-similar prescaling indicative of a nonthermal attractor.
    This behavior can be analytically understood through perturbation theory for the Polyakov loop correlations and the established dynamics of gauge field correlations.
    We perform simulations for both $\SU(2)$ and $\SU(3)$ gauge groups, providing further consistency checks on the $N_c$-dependence of perturbative expectations.
    No evidence of an inverse cascade toward lower momenta is found for momenta above the electric screening scale.
\end{abstract}

\maketitle

\section{Introduction}

Understanding the dynamics of strongly correlated nuclear matter, especially in high-energy heavy-ion collisions, poses a significant challenge to theoretical approaches based on first principles~\cite{Busza:2018rrf, Schlichting:2019abc, Berges:2020fwq}.
The quark-gluon plasma formed within the first yoctosecond of the collision leads to phenomena such as gluon saturation and over-occupation~\cite{Gribov:1983ivg, Gelis:2010nm}.
The subsequent evolution of the gluon-dominated plasma often exhibits universal dynamics, including universal time dependencies related to nonthermal fixed points~\cite{Berges:2013eia, Berges:2013fga, Berges:2014bba, Berges:2008wm}, as well as hydrodynamic attractors~\cite{Heller:2015dha, Romatschke:2017vte, Soloviev:2021lhs}. 

Nonthermal fixed points occur far from equilibrium and are marked by self-similarity in the time evolution of correlations, characterized by universal scaling exponents and a universal shape of the distribution function within a specific momentum region.
This allows systems to be grouped into far-from-equilibrium universality classes.
In particular, relativistic and nonrelativistic scalars form one such class at low momenta based on equal-time distributions~\cite{PineiroOrioli:2015cpb}, which can be  refined further based on unequal-time and topological observables~\cite{Boguslavski:2019ecc, Noel:2023oyz}. 
For longitudinally expanding systems, non-Abelian gauge fields and scalars constitute another class at high momenta~\cite{Berges:2014bba}.
Crucially, scalar $\mathrm{O}(N)$ models exhibit a dual cascade near a nonthermal fixed point, which has been observed in a variety of experiments with cold quantum gases~\cite{Prufer:2018hto, Erne:2018gmz, Glidden:2020qmu, Gazo:2023exc}. 
Here, energy is transported toward the ultraviolet in a direct cascade, while particle number is transported toward the infrared in an inverse cascade, dynamically generating a scalar condensate in the zero mode of the distribution function~\cite{Berges:2012us, PineiroOrioli:2015cpb}.

In the case of gluons, while a direct cascade is observed, to date there is no evidence of infrared dynamics that would lead to low-momentum correlations or condensate formation in gauge-fixed gluon distribution functions in terms of gauge field correlations~\cite{Berges:2008mr, Kurkela:2011ti, Kurkela:2012hp, Berges:2012ev, Schlichting:2012es, Berges:2013fga, AbraaoYork:2014hbk, Blaizot:2016iir, Mace:2016svc, Boguslavski:2018beu, Boguslavski:2019fsb} that had been previously proposed to emerge in over-occupied gluonic plasmas \cite{Blaizot:2011xf, Blaizot:2013lga}. 
Defining a proper gauge-invariant condensate in gauge theory is, however, nontrivial.
Recently, it has been found that out-of-equilibrium gluonic matter can generate a gauge-invariant condensate defined in terms of nonlocal order parameters, such as spatial Wilson loops or Polyakov loop correlations~\cite{Berges:2019oun, Berges:2023sbs}.
While Wilson loops have been used previously in studying topological transitions and string tension dynamics out of equilibrium~\cite{Mace:2016svc, Berges:2017igc}, Polyakov loop traces provide a gauge-invariant (scalar) field whose correlations exhibit the emergence of a sizable condensate fraction. Furthermore, they can be used to formulate a low-energy effective action~\cite{Berges:2023sbs}.
Related gauge-invariant operators have also been used in contexts like the Abelian Higgs model and non-Abelian gauge theories on continuous toroidal space-times~\cite{Mitreuter:1996ze, Ford:1998bt, Gasenzer:2013era}.

\begin{figure}[t]
    \centering
    \includegraphics[scale=0.85]{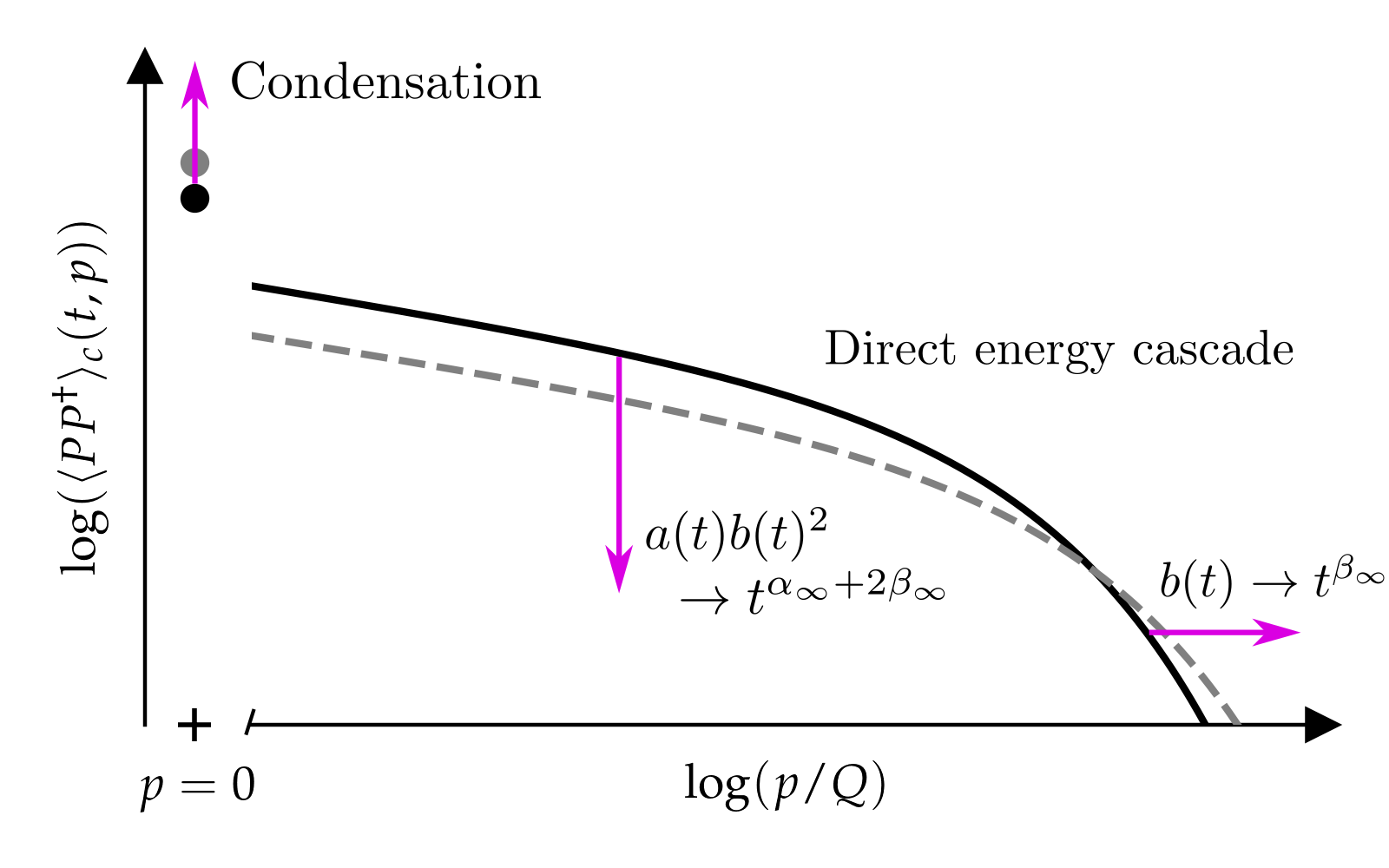}
    \caption{Schematic picture of the evolution of Polyakov loop correlations, highlighting the build-up of a condensate in the zero mode and a direct cascade toward the ultraviolet exhibiting prescaling dynamics.
    There are no indications for an inverse cascade toward lower momenta in the investigated momentum regime.
    Pink arrows indicate dynamics for growing times.
    The prescaling factors $a(t)$, $b(t)$ and asymptotic scaling exponents $\alpha_\infty,\beta_\infty <0$ are defined for gluon momentum distribution functions.
 }\label{Fig:PPCorrelationSchematics}
\end{figure}

In this work, we extend the study of Polyakov loop correlations to momenta above the electric screening scale.
Our study is based on real-time $\SU(N_c)$ lattice gauge theory simulations in the classical-statistical regime of weak couplings and large gluon occupations, which describes well the early-time evolution of the quark-gluon plasma~\cite{Gelis:2010nm, Kurkela:2012hp, Berges:2013eia, Skullerud:2003ki, Mueller:2002gd, Jeon:2004dh, AbraaoYork:2014hbk, Aarts:2001yn, Arrizabalaga:2004iw}. 
We carry out simulations for gauge groups $\SU(2)$ and $\SU(3)$ and find a relatively simple composition of the dynamics, as sketched in \Cref{Fig:PPCorrelationSchematics}.
It features the growth of the condensate fraction from early times, consistent with previous findings~\cite{Berges:2023sbs}, while the (Polyakov loop correlator) distribution together with the energy of hard gluons is transported toward the ultraviolet in a direct, self-similar energy cascade.
We find no evidence of an inverse cascade at low momenta that would populate the zero mode.

The description of the direct energy cascade requires time-dependent scaling exponents, which only asymptotically acquire their universal values. While these late-time values coincide with the universal scaling exponents of the original gluon distributions, their time-dependent deviations are a sign of prescaling.
For a kinetic theory of quantum chromodynamics (QCD)~\cite{Mazeliauskas:2018yef} and for simulations of a 3-dimensional Bose gas~\cite{Schmied:2018upn}, prescaling regimes have been identified before. 
In the context of \cite{Mazeliauskas:2018yef} that we follow in our work, prescaling occurs when the shape of correlations becomes universal much faster than the scaling exponents reach their universal values. 
It has been shown that prescaling provides a generic phenomenon characteristic of the approach to far-from-equilibrium attractors~\cite{Heller:2023mah}, which has subsequently also been found experimentally~\cite{Gazo:2023exc}.

Our numerical results are complemented by perturbative predictions for the Polyakov loop correlations, which provide analytic access to the asymptotic scaling behavior near the nonthermal attractor. 
We demonstrate that they agree well with our simulations and are based on the known behavior of gauge field correlations in this regime.
Analytically, gauge invariance of the perturbative Polyakov loop correlations is preserved through the appearance of gauge transformation averages.
As an additional validation, our simulations align with the nontrivial perturbative $N_c$-dependence of Polyakov loop correlations.

This paper is structured as follows: \Cref{Sec:LatticeSetup} introduces the employed lattice setup.
Subsequently, \Cref{Sec:Correlations} presents results for the Polyakov loop correlations at zero and nonzero momenta.
We discuss the formation of a condensate and the relationship between ultraviolet dynamics, energy transport, and prescaling.
The section concludes with an analysis of the $N_c$-dependence of the correlations.
\Cref{Sec:Analytics} provides an analytic, perturbative derivation of the asymptotic scaling behavior of the Polyakov loop correlations and their $N_c$-dependence, based on the known dynamics of gauge field correlations.
Finally, in \Cref{Sec:Conclusions}, we conclude and give an outlook.

\section{Numerical lattice setup}\label{Sec:LatticeSetup}

We consider $\SU(N_c)$ lattice gauge theories in 3+1 space-time dimensions with the Minkowski metric for $N_c=2$ and $N_c=3$.
Fields are discretized on a cubic spatial lattice with $N_s^3$ lattice sites for spatial lattice spacing $a_s$, so that the physical volume of the spatial lattice $\Lambda_s=\{1,\ldots,N_s\}^3$ is $V=L_s^3=a_s^3 N_s^3$. 
Spatially periodic boundary conditions are employed and the time evolution of the theory is simulated in time steps of width~$\dd t$.
We work in temporal-axial gauge, $A_0\equiv 0$.
The theory is formulated on the lattice in terms of $\su(N_c)$-valued \mbox{(chromo-)}electric fields $E_i(t,\xx)$ and, to preserve gauge covariance, $\SU(N_c)$-valued link variables $U_i(t,\xx)$, with $i=1,2,3$, $\xx=(x_1,x_2,x_3)$.
The latter are related to the gauge fields $A_i(t,\xx)$ via $U_i(t,\xx)\approx \exp(-i g a_s A_i(t,\xx))$ up to lattice corrections, where $g$ is the gauge coupling.

In this work, we employ highly occupied gluon initial conditions at weak couplings $g\ll 1$. 
At sufficiently early times, gluonic occupation numbers remain large, and the full quantum dynamics can be approximated well by a classical-statistical evolution~\cite{Aarts:2001yn, Mueller:2002gd, Skullerud:2003ki, Jeon:2004dh, Arrizabalaga:2004iw, Kurkela:2012hp, Berges:2013lsa}. 
Specifically, we sample over Gaussian initial conditions with variances
\begin{subequations}
\label{eq:init_cond}
\begin{align}
\langle AA\rangle (t{=}0,\pp) = &\; \frac{Q}{g^2 |\pp|^2}\, \theta(Q-|\pp|)\,,\\
\langle EE\rangle(t{=}0,\pp) = &\; \frac{Q}{g^2}\, \theta(Q-|\pp|)\,,
\end{align}
\end{subequations}
where averages over transverse polarizations and color degrees are implied, and the momentum scale $Q$ determines the width of the initial momentum space distribution.
Such gluon over-occupation up to a gluon saturation scale is characteristic of a state shortly after the collision of heavy nuclei~\cite{Berges:2020fwq}.
Gauss' law is fixed initially by means of a projection algorithm onto the constraint surface~\cite{Moore:1996qs} and is preserved by the time evolution.
Given the initial conditions, in the classical-statistical approximation, the time evolution of fields is computed for each initial configuration separately through the equations of motion that follow from the classical lattice Hamiltonian
\begin{align}\label{EqHamiltonian}
H(t)=&\; \frac{a_s^3}{g^2}\sum_{\xx\in\Lambda_s}\bigg[ \Tr(\EE(t,\xx)^2)\nonumber\\
&\qquad + \frac{2}{a_s^4}\sum_{j>k}[N_c - \Re{\Tr(U_{jk}(t,\xx))}]\bigg]\,,
\end{align}
with the lattice plaquette variables
\begin{equation}
U_{jk}(t,\xx) = U_j(t,\xx)U_k(t,\xx + \jhat) U_j^{\dagger}(t,\xx+ \khat) U_k^\dagger(t,\xx)\,.
\end{equation}
Observables are then calculated by averaging over the resulting independent trajectories at the time of interest. 
More details regarding the initial conditions and the lattice setup can be found, e.g., in~\cite{Boguslavski:2018beu, Berges:2013fga}.

For our simulations, we choose $N_c=2$ if not stated otherwise and ${N_c=3}$ for comparison, and set $N_s=256$, $\dd t / a_s = 0.05$, and $Qa_s = 0.125$.
We consider fields rescaled as $A\mapsto g A$ so that, with our initial conditions \eqref{eq:init_cond}, the formulation becomes independent of the coupling, which is a property of the classical-statistical framework.
We explicitly verified the insensitivity of our correlations at finite momentum to variations of the lattice parameters, specifically upon comparison with results for $Q a_s = 0.25$ with the same lattice volume and with results for a larger spatial lattice size $N_s=512$.
If not stated otherwise, the shown results are averaged over five simulations.

\section{Results for Polyakov loop correlations}\label{Sec:Correlations}

\subsection{Correlations of spatial Polyakov loop traces}\label{Sec:PolyakovCorrelations}

\begin{figure}[t]
    \centering
	\includegraphics[scale=0.85]{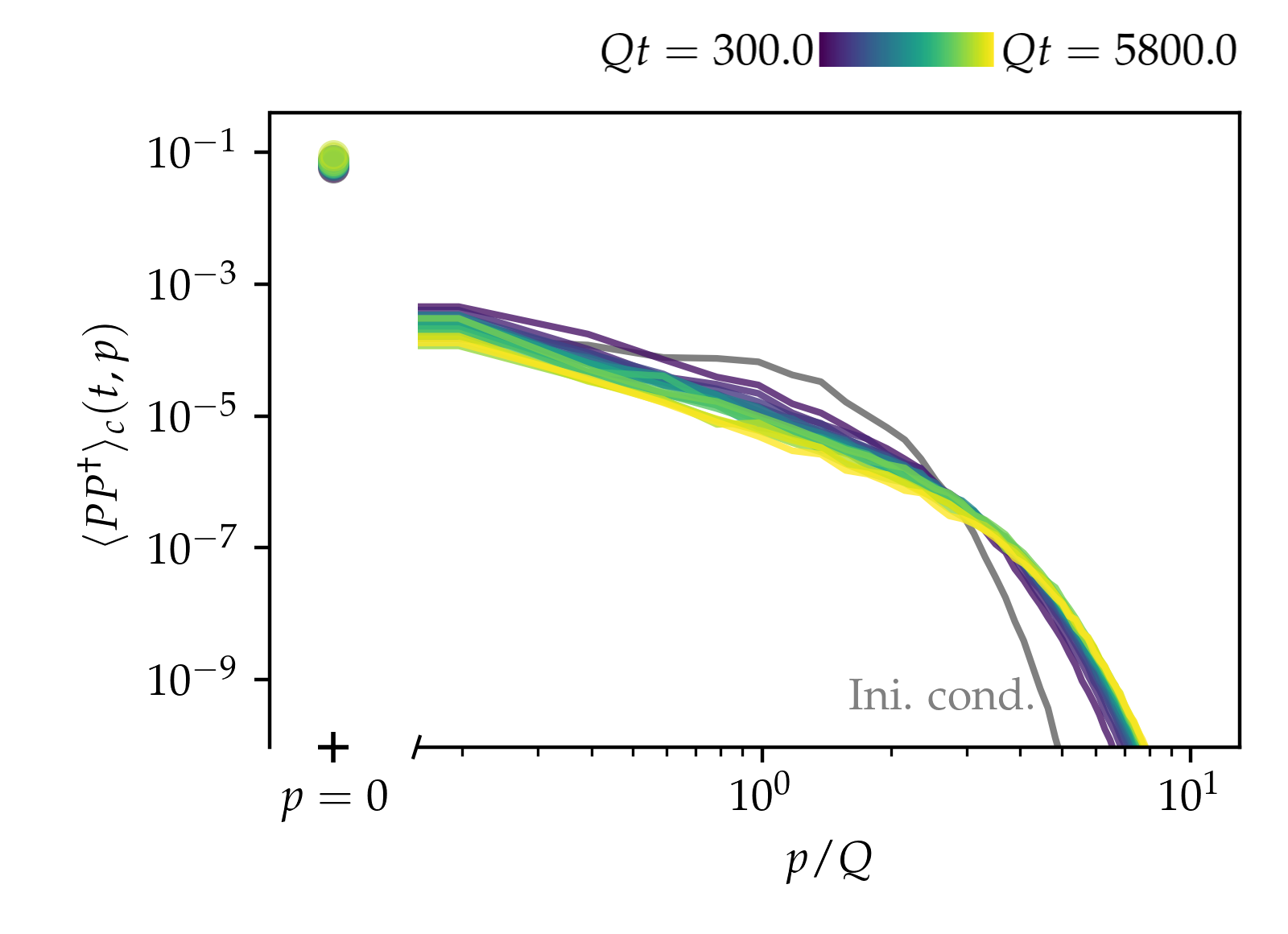}
	\caption{Correlations $\langle P P^\dagger\rangle_c(t,p)$ versus momentum $p$ for gauge group $\SU(2)$, shown for different times as indicated by the colorbar.
    Absolute physical momenta in the 1-3 plane are employed: $p=|\pb|$.
    For the zero momentum correlations, the data at five consecutive readout times is averaged as detailed in the caption of \Cref{Fig:PolyakovLoopCorrelationsModeShare}.
 }\label{Fig:PolyakovLoopCorrelations}
\end{figure}

We define the spatial Polyakov loop trace as
\begin{equation}
    P(t,\xb):=\frac{1}{N_c}\Tr \,\Pscr \prod_{x_2=1}^{N_s} U_2(t,\xx)\,,
\end{equation}
where $\xb=(x_1,x_3)$ and $\Pscr$ indicates path ordering along the positive 2-direction.
Note that $P(t,\xb)$ is real-valued for $N_c=2$ but complex-valued for $N_c=3$.%
\footnote{We do not restrict to the real part of $P(t,\xb)$ here in order to facilitate the perturbative description detailed in \Cref{Sec:Analytics}, see in particular \Cref{Sec:AnalyticsPolyakovCorrelationsFromGauge}.} 
It is gauge-invariant by construction and has only (2+1)-dimensional space-time support since it describes a Wilson loop trace winding around the periodic 2-direction.

We focus on connected 2-point correlations of the spatial Polyakov loop traces as provided by
\begin{align}
    &\langle P(t,\xb) P^\dagger (t,\yb)\rangle_c \nonumber\\
    &\qquad:=\; \langle P(t,\xb) P^\dagger (t,\yb)\rangle - \langle P(t,\xb)\rangle \langle P^\dagger (t,\yb)\rangle\,.
\end{align}
In spatially homogeneous settings, i.e., where the system does not depend on $\xb+\yb$ such as ours, these correlations are suitably described in momentum space:
\begin{equation}\label{eq:PPcorrelatorpspace}
    \langle PP^\dagger \rangle_c (t,\pb):= a_s^2 \sum_{\Delta\xb\in \bar{\Lambda}_s} \langle P(t,\Delta\xb) P^\dagger (t,\zerob)\rangle_c \, e^{-i \ptilde\Delta\xb}\,,
\end{equation}
where $\pb = (p_1,p_3)$ is the physical momentum and $\ptilde=(\tilde{p}_1,\tilde{p}_3)$ is the lattice momentum, $\tilde{p}_i\in 2\pi/(a_s N_s)\{ -N_s/2,-N_s/2+1,\ldots,N_s/2-1 \}$.
The physical momenta $p_i$ are related to the lattice momenta via $p_i=(2/a_s)\sin(\tilde{p}_i a_s/2)$.
For configurations, which are furthermore spatially isotropic in the 1-3 plane, the correlations~\eqref{eq:PPcorrelatorpspace} reduce to $\langle PP^\dagger \rangle_c (t,p) = \langle PP^\dagger \rangle_c (t,\pb)$ for $p=|\pb|$ and are the main observables of interest in this work.

In \Cref{Fig:PolyakovLoopCorrelations} we display the Polyakov loop correlations $\langle PP^\dagger\rangle_c(t,p)$ for both the zero mode ($p=0$) and the nonzero modes ($p>0$) as a function of time, including the initial time.
We find that the $p=0$ correlation grows with time, whose relation to the dynamic formation of a gauge-invariant condensate in terms of Polyakov loops will be discussed in more detail in \Cref{Sec:CondensateFraction}.
For momenta $p>0$, initial correlations spread at early times toward the ultraviolet.
After some time, the shape of correlations becomes universal, i.e., remains approximately invariant under time evolution, which is the temporal regime that we focus on.
In this regime, the correlations decrease in overall value and shift further toward larger momenta with increasing time.
For momenta $p/Q\lesssim 2.0$, the shape of the correlations can be well approximated by a power law with an exponent between -2.0 and -1.5 (not displayed).
In terms of overall numbers, the zero mode correlation is much larger than those at individual nonzero momenta.
Crucially, based on the Polyakov loop correlations analyzed, no inverse cascade---like those seen in scalar theories under weakly coupled, over-occupied conditions~\cite{Berges:2012us, PineiroOrioli:2015cpb}---is observed within the investigated momentum region.

Interestingly, while it was proposed that a large initial density of gluons could facilitate the formation of a condensate \cite{Blaizot:2011xf, Blaizot:2013lga}, for the gluon distribution function below the Debye scale $m_D$, neither condensate formation nor an inverse cascade has been observed, which can be attributed to a nontrivial cancellation of elastic and inelastic processes at low momenta~\cite{Kurkela:2012hp, Berges:2013eia, AbraaoYork:2014hbk, Blaizot:2016iir}.
However, the perturbative relation between the gluon distribution function and the Polyakov loop correlator, which we derive in \Cref{eq:Polyakov2ptPerturb} below, does not apply at low momenta $p \lesssim m_D$, where occupation numbers can become large. Thus, we cannot exclude the presence of an inverse cascade in Polyakov loop correlations below the electric screening scale $m_D$.
In our simulations, this scale decreases from $m_D/Q=0.23$ at time $Qt=300$ to $m_D/Q=0.14$ at time $Qt=5800$, and is therefore of the order of the smallest nonzero momenta resolved on our lattice.%
\footnote{The Debye mass can be computed from the transversely polarized gauge field correlator $\langle AA\rangle(t,p)$ according to $m_D^2(t) = 4 N_c \int \dd^3 p/(2\pi)^3 \langle AA\rangle(t,p)$~\cite{Boguslavski:2018beu}.}

\subsection{Infrared: dynamic condensate formation}\label{Sec:CondensateFraction}

\begin{figure}[t]
    \centering
	\includegraphics[scale=0.85]{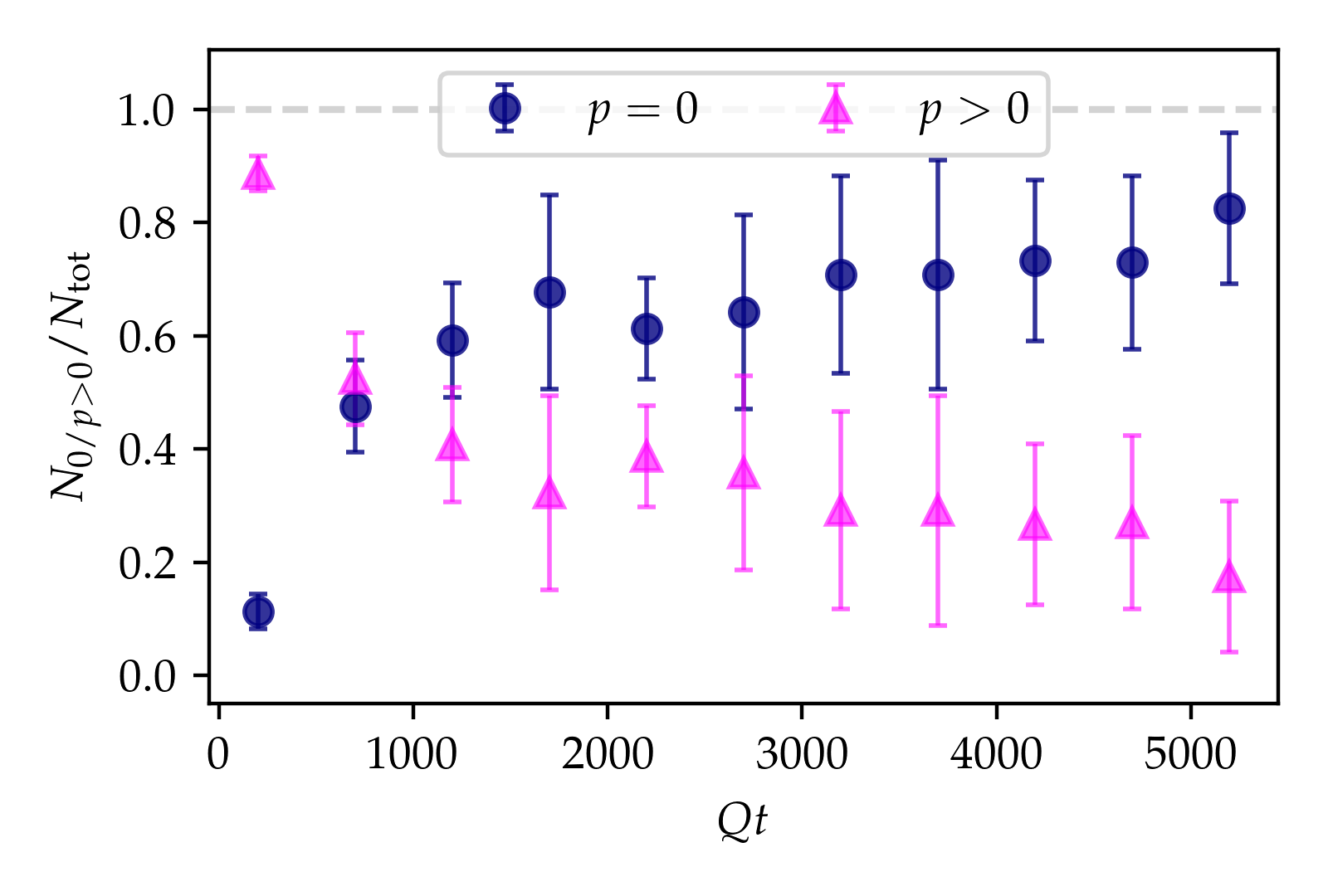}
	\caption{The share of Polyakov loop correlations in the zero mode (condensate fraction $N_0(t)/N_\tot(t)$, blue circles) versus the sum of higher modes ($N_{p>0}(t)/N_\tot(t)$, pink triangles) as functions of time for an $\SU(2)$ plasma. 
    The evolution highlights the build-up of a condensate through the redistribution of Polyakov loop correlations to the zero mode.
    Error bars are computed as standard errors of the mean over five simulation runs. 
    Additionally, data points at time $Qt$ are averaged over the readout times $Qt-200$, $Qt-100$, $Qt$, $Qt+100$, ${Qt+200}$ and the resulting uncertainty is computed via error propagation.
 }\label{Fig:PolyakovLoopCorrelationsModeShare}
\end{figure}

In order to further study the behavior of the system with regard to the zero mode of the Polyakov loop correlations, we consider the contributions of both the $p=0$ mode and the $p>0$ modes to the sum of all correlations.
We define zero and nonzero mode Polyakov loop occupations:
\begin{subequations}
    \begin{align}
        N_0(t) := &\; \frac{1}{(N_s a_s)^2}\langle PP^\dagger\rangle_c(t,\pb=\zerob)\,,\\
        N_{p>0}(t) := &\; \frac{1}{(N_s a_s)^2}\sum_{\pb\neq \zerob}\langle PP^\dagger\rangle_c(t,\pb)\,,
    \end{align}
\end{subequations}
along with their sum $N_\tot(t):=N_0(t)+N_{p>0}(t)$.
The sum in $N_{p>0}(t)$ and its prefactor correspond to a discretized version of $\int d^2p/(2\pi)^2$.
In \Cref{Fig:PolyakovLoopCorrelationsModeShare} we show the two quotients $N_0(t)/N_\tot(t)$ and $N_{p>0}(t)/N_\tot(t)$ as functions of time.
The variable $N_0(t)/N_\tot(t)$ is analogous to the condensate fraction of excitations in the local Polyakov loop traces~\cite{Berges:2023sbs} and measures the amount of long-range coherence in the system~\cite{Glauber:1963fi}.

At the initial time $t=0$, the total amount of correlations is strongly dominated by nonzero momenta and the zero mode barely contributes, $N_{p>0}(0)/N_\tot(0) \gg N_0(0)/N_\tot(0)$.
At early times, $N_0(t)/N_\tot(t)$ starts to build up, indicating the dynamic formation of a Polyakov loop condensate.
Simultaneously, the contribution of nonzero momentum modes to the total correlations steadily diminishes, though this trend gradually slows down over time.

To confirm the presence of a condensate, one would need to analyze the volume scaling of the time required for condensate formation.
This was previously demonstrated for similar initial conditions in \cite{Berges:2019oun, Berges:2023sbs}, where the condensate formation time was numerically shown to scale with an approximately universal power of volume, and the condensate build-up followed a universal, volume-independent form.
We note that the Polyakov-loop-based condensate fraction defined in~\cite{Berges:2023sbs} from position space correlations is almost identical to $N_0/N_\tot$ as considered in this work.

\subsection{Ultraviolet: energy transport in a prescaling cascade}\label{Sec:PrescalingUV}

\begin{figure*}[t]
    \centering
	\includegraphics[scale=0.85]{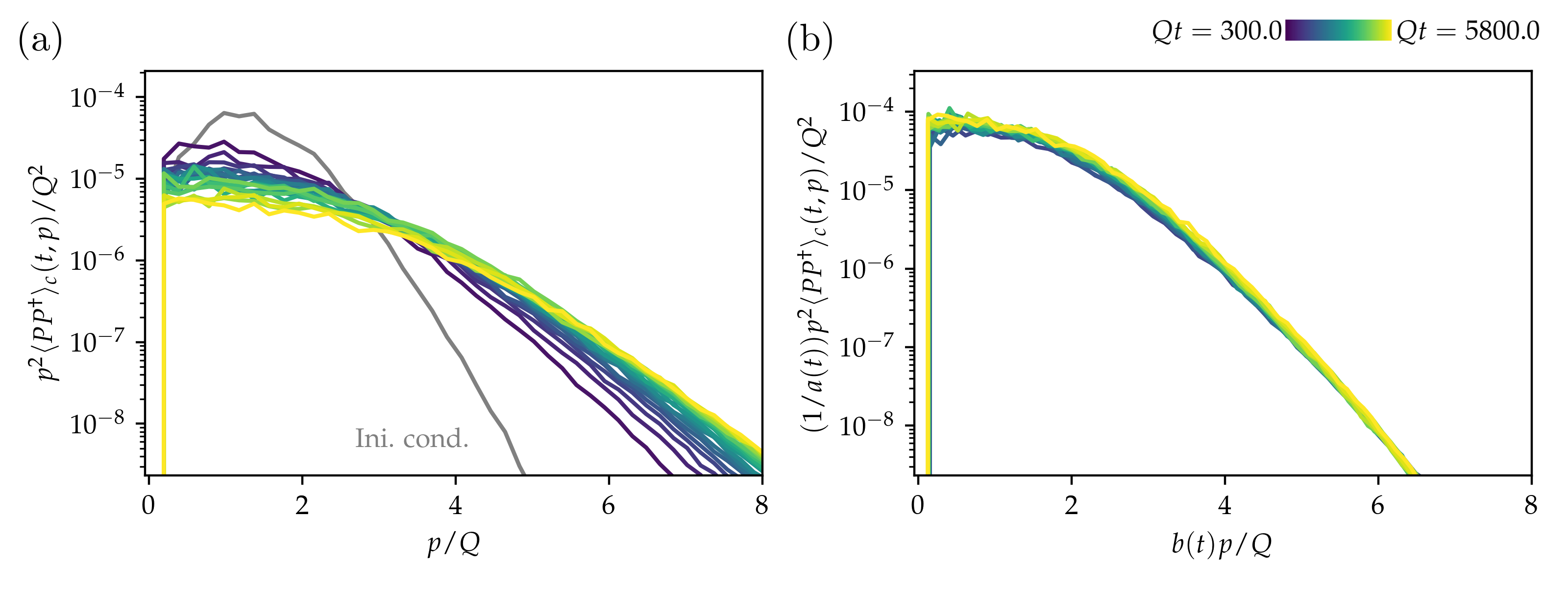}
	\caption{(a)~Correlations $p^2 \langle P P^\dagger\rangle_c(t,p)$ versus momentum $p$ shown for different times as indicated by the colorbar.
 (b)~Corresponding rescaled correlations for the prescaling functions $a(t),b(t)$ extracted as detailed in \Cref{Sec:PrescalingUV}.
 The prefactor for the temporal scaling with the prescaling function $a(t)$ follows from \Cref{eq:prescalingPP} upon inclusion of the $p^2$ factor.
 Here, absolute physical momenta in the 1-3 plane are employed: $p=|\pb|$.
}
\label{Fig:PolyakovLoopCorrelationsRescaled}
\end{figure*}

We turn to the dynamics of the correlations for nonzero momenta in more detail.
As shown in \Cref{Fig:PolyakovLoopCorrelations}, the shape of $\langle PP^\dagger\rangle_c(t,p)$ stays constant in time, up to statistical fluctuations.
This can be characteristic of the dynamics of a self-similar cascade near a nonthermal fixed point. Self-similar scaling emerges already in the approach to such an attractor~\cite{Micha:2004bv, Mazeliauskas:2018yef, Schmied:2018upn, Heller:2023mah}, a phenomenon called prescaling~\cite{Mazeliauskas:2018yef}. 
Prescaling has been investigated so far only through distribution functions such as that of gluons~\cite{Mazeliauskas:2018yef, Heller:2023mah}, which is, up to momentum-dependent prefactors, proportional to the gauge-dependent gluon-gluon correlations $\langle A A\rangle (t,p)$.
Specifically, near a nonthermal attractor, their time dependence can be suitably described by%
\footnote{Let $f(t,p)$ denote the gluon distribution function, which for momenta far above the thermal mass is related to transversely polarized gluon correlations via $f(t,p)\sim p \, \langle AA\rangle_c(t,p)$.
The prescaling behavior $f(t,p)=a(t) f_S(b(t) p)$ then implies $\langle AA\rangle_c(t,p) = a(t) b(t) \langle AA\rangle_{c,S}(b(t) p)$, where $f_S(p)$ and $\langle AA\rangle_{c,S}(p)$ respectively denote the fixed-point shapes of $f(t,p)$ and $\langle AA\rangle_c(t,p)$.
}
\begin{equation}\label{Eq:PrescalingAnsatzAA}
    \langle AA\rangle(t,p) = a(t)b(t)\, \langle AA\rangle_S(b(t)p)\,.
\end{equation}
Here, $\langle AA\rangle_S(p)$ indicates the fixed-point shape of $\langle AA\rangle(t,p)$, and $a(t)$, $b(t)$ denote prescaling factors, which can equivalently be used to introduce time-dependent scaling exponents as $a(t)=\exp(\int_{t_0}^t \dd t'\, \alpha(t')/t')$ and ${b(t)=\exp(\int_{t_0}^t \dd t'\, \beta(t')/t')}$.
The parameter $t_0$ specifies an initial time scale for the prescaling dynamics.
For large times, if a nonthermal attractor is approached, $\alpha(t')$ and~$\beta(t')$ converge to their asymptotic values $\alpha_\infty$ and~$\beta_\infty$.
In fact, much research on nonthermal fixed points has been based on these universal asymptotic values~\cite{Berges:2020fwq}, but it has been shown in kinetic theory that prescaling appears generally in the vicinity of nonthermal attractors~\cite{Heller:2023mah}.
Usually, the instantaneous preservation of emergent conservation laws for the distribution function provides a link between $\alpha(t')$ and $\beta(t')$ during their evolution.

The prescaling behavior of the gluon correlations can be related to the Polyakov loop correlations by means of the leading-order perturbative expression 
\begin{align}
    &\langle PP^\dagger\rangle_c(t,p) = \frac{g^2}{N_c}\int \Dscr \Gcal \int \dd^2 (\xb-\yb)\oint_0^{L_s}\dd x_2  \oint_0^{L_s} \dd y_2 \nonumber\\
    &\qquad\quad \times \Tr\, \langle A_2(t,\xx) A_2(t,\yy)\rangle_c\, e^{-i\pb (\xb-\yb)} + \Ocal(g^3)\,,
\end{align}
that we will derive for a continuous space-time with spatially periodic boundary conditions in \Cref{Sec:Analytics}, to which we also refer for notational details.
Therefore, if the system shows prescaling in the gauge field correlator and the perturbative treatment is justified, the Polyakov loop correlations may inherit this property. 
This motivates the ansatz
\begin{equation}\label{eq:prescalingPP}
    \langle PP^\dagger\rangle_c(t,p) = a(t) b^2(t) \langle PP^\dagger\rangle_{c,S}(b(t)p)\,,
\end{equation}
where $\langle PP^\dagger\rangle_{c,S}(p)$ indicates the fixed-point shape of $\langle PP^\dagger\rangle_c(t,p)$.

Asymptotically, energy transport toward the ultraviolet for gluonic plasmas can proceed in a self-similar cascade with $\beta_\infty=-1/7$, which has been observed in the gluon distribution function~\cite{Schlichting:2012es, Kurkela:2012hp, Berges:2013fga, AbraaoYork:2014hbk, Boguslavski:2018beu} and in energy density correlations~\cite{Spitz:2023wmn}.
In physical terms, this implies that the hard scale that dominates the energy density grows with time asymptotically as $\sim Q (Qt)^{1/7}$.
For relativistic gluons in 3 spatial dimensions, energy conservation yields ${\alpha(t')=4\beta(t')}$, which generalizes the well-known relation for their asymptotic values ($\alpha_\infty = 4\beta_\infty$).

In order to investigate to what extent the ansatz \eqref{eq:prescalingPP} describes the Polyakov loop correlations well, in \Cref{Fig:PolyakovLoopCorrelationsRescaled}(a) we show $p^2 \langle PP^\dagger\rangle_c(t,p)$ for sufficiently late times when the correlations approximately maintain their shape.
The depicted second moment of the correlations has proven particularly sensitive to self-similarity and thus suitable to visually verify the presence of prescaling~\cite{Heller:2023mah}.
Importantly, there are scaling functions $a(t)$ and $b(t)$ through which we can rescale this quantity so that the curves lie on top of each other up to numerical fluctuations for the shown time interval, see \Cref{Fig:PolyakovLoopCorrelationsRescaled}(b).
This reveals the presence of a prescaling cascade in the Polyakov loop correlations.

\begin{figure}[t]
    \centering
	\includegraphics[scale=0.85]{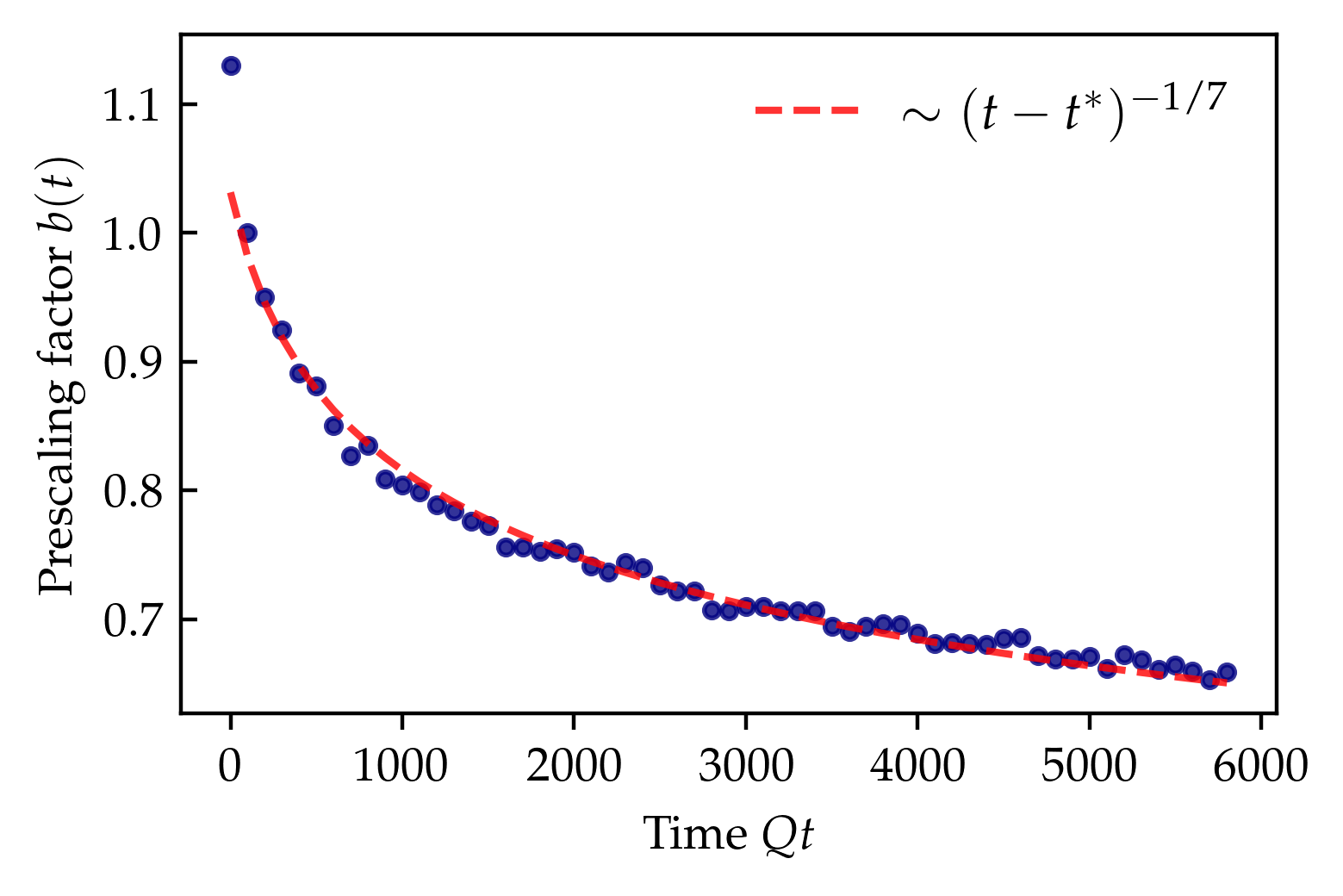}
	\caption{The numerically extracted prescaling factor $b(t)$ shown as a function of time, similar to~\cite{Heller:2023mah}. 
 Prescaling factors extracted from different moments are averaged over, see \Cref{appendix:Prescaling}. 
 The dashed red line corresponds to the prescaling solution in \Cref{eq:prescaling_b}.
}\label{Fig:UVPrescalingExponents}
\end{figure}

The prescaling factors $a(t),b(t)$ are extracted from different moments of the Polyakov loop correlations, reflecting that different momentum scales show the same scaling behavior in time.
The time evolution of $b(t)$, in particular, is shown in~\Cref{Fig:UVPrescalingExponents} to be described well by the prescaling solution as derived in~\cite{Heller:2023mah}:
\begin{align} \label{eq:prescaling_b}
 b(t)=((t-t_*)/t_{\mathrm{ref}})^{\beta_\infty},
\end{align}
with observable- and initial condition-dependent parameters $t_*$ and $t_{\mathrm{ref}}$ and with the universal exponent $\beta_\infty$. 
The time scale $t_*$ is fixed by data at time $t_0$ and $t_{\mathrm{ref}}$ can be viewed as a normalization of the gluon distribution function.
The function $a(t)$ numerically obeys $a(t) = b(t)^4$, or equivalently $\alpha(t)=4\beta(t)$, in agreement with energy conservation, see Fig.~\ref{Fig:AppEnergyConservation} in the Appendix. 
The extraction procedure and details on prescaling are described in \Cref{appendix:Prescaling}.

\begin{figure*}
    \centering
	\includegraphics[scale=0.85]{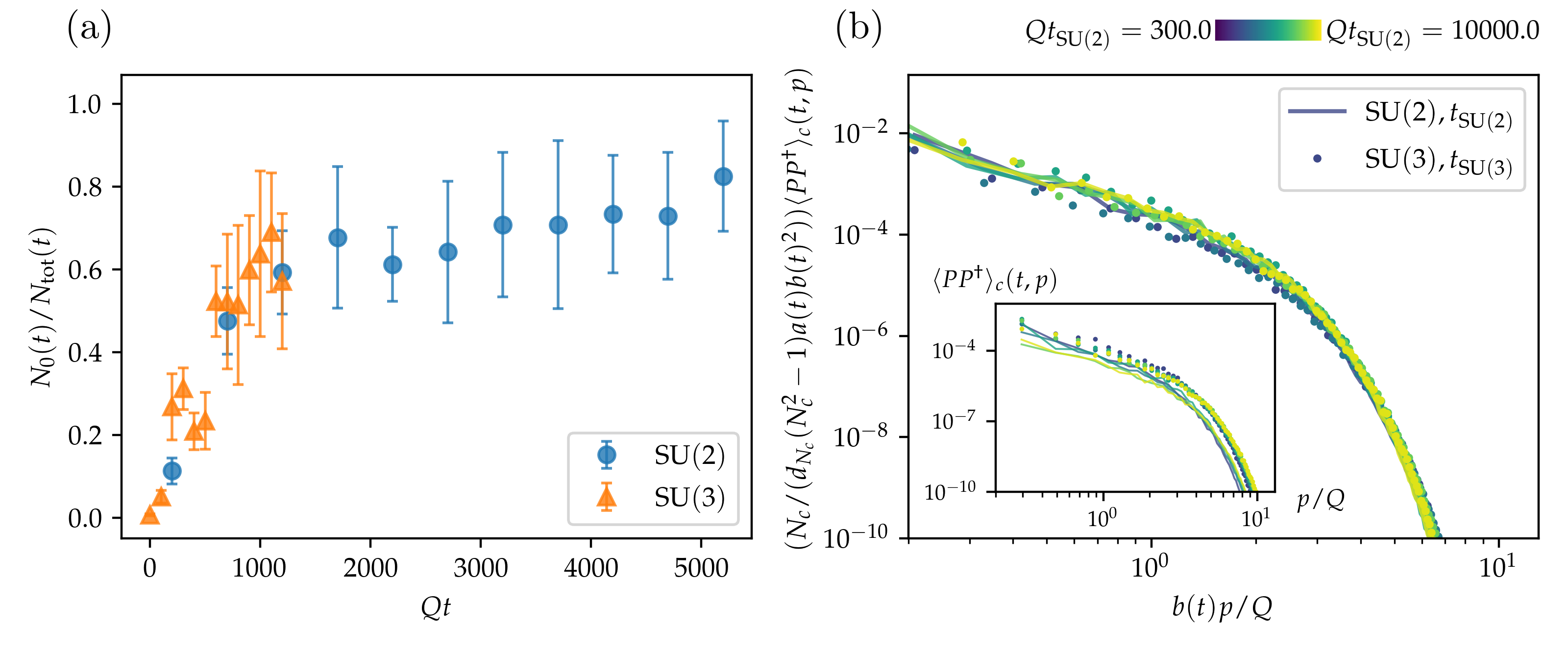}
	\caption{$N_c$-dependence of the Polyakov loop correlations.
 (a)~Condensate fraction as a function of time for gauge groups $\SU(2)$ (blue disks) and $\SU(3)$ (orange triangles).
 Error bars are computed as averages over five simulation runs. For $\SU(2)$, the additional time average as detailed in the caption of \Cref{Fig:PolyakovLoopCorrelationsModeShare} is used.
 (b)~Connected Polyakov loop correlations versus physical momentum $p=|\pb|$, shown for a range of times as indicated by the colorbar for the $\SU(2)$ case (solid lines; provided are $Qt_{\SU(2)} = 2200, 4000, 5800, 7600, 9400$).
 In the case of gauge group $\SU(3)$, the corresponding times are given by $t_{\SU(3)} = (2/3)^2 t_{\SU(2)}$ in accordance with the considerations in the main text (points; displayed are $Qt_{\SU(3)} = 1000,1800,2600,3400,4200$).
 For both gauge groups, the employed prescaling factors are those for the times $Q t_{\SU(2)}$.
 The inset in (b) shows the results without $N_c$- and time rescalings between $\SU(2)$ and $\SU(3)$; the displayed times in the inset of (b) are the ones for $\SU(3)$.
 The results shown in (b) are consistent with the perturbative analytics provided in \Cref{Sec:Analytics}.
 They originate from a single simulation run due to longer available simulation times.
 }\label{Fig:NcDependence}
\end{figure*}

\subsection{Dependence on $N_c$}
\label{Sec:NcDependence}

We finally investigate the dependence of the Polyakov loop correlations on the number of colors $N_c$.
We start with the condensate fraction that is shown in \Cref{Fig:NcDependence}(a) for the gauge groups $\SU(2)$ and $\SU(3)$ as functions of time for the same initial conditions.
The behavior of the condensate fraction computed for the gauge group $\SU(3)$ is very similar to $\SU(2)$ and the considerations of \Cref{Sec:CondensateFraction} apply \emph{ad verbatim}.
Within numerical uncertainties, both dynamics are comparably fast and no $N_c$-dependence is visible with the current level of accuracy.
Due to the relatively large error bars, a decisive study on a possible $N_c$-dependence of the condensate growth in Polyakov loop correlations would require much better statistics and is beyond the scope of the present work.

In order to understand the behavior of the Polyakov loop correlations at nonzero momenta, we first examine a potential non-trivial $N_c$-dependence of the general speed of the dynamics, which is based on the kinetic theory arguments outlined in~\cite{AbraaoYork:2014hbk}.
For the times under investigation, the system can be well described within QCD effective kinetic theory \cite{Arnold:2002zm} by means of a Boltzmann equation for the gluon distribution function~$f(t,p)$:
\begin{equation}\label{eq:Boltzmann}
    \frac{\dd f(t,p)}{\dd t} = -\Ccal[f]\,.
\end{equation}
The collision kernel $\Ccal[f]$ consists of a sum of elastic and inelastic collision kernels, which we can consider for large gluon occupancies in the classical approximation. 
The leading $N_c$ scaling comes from the elastic collision kernel $\Ccal[f]\sim N_c^2$ so that \Cref{eq:Boltzmann} becomes approximately $N_c$-independent if we consider all times rescaled as $N_c^2 t$ instead of $t$.
The remaining $N_c$ dependence lies in the inelastic collision kernel and in the soft screening scale~$m_D$~\cite{AbraaoYork:2014hbk}, which the numerical results seem not to be sensitive to at high momenta.

Based on this scaling argument, to properly compare the Polyakov loop correlations at nonzero momenta across different $N_c$ for the same initial condition, we need to consider different times for the gauge groups $\SU(2)$ and $\SU(3)$.
In \Cref{Fig:NcDependence}(b) we show the correlations for the two gauge groups for times $Qt_{\SU(2)} = 2200,\ldots,9400$ and $Qt_{\SU(3)}=1000,\ldots,4200$, all well within the prescaling regime discussed before.
The inset shows the correlations for the times $Qt_{\SU(3)}$ also in the $\SU(2)$ case and without any $N_c$-dependent rescalings.
Clearly, the correlations exhibit a nontrivial $N_c$-dependence, even though the shape of the $\SU(3)$ results is comparable to that of the gauge group $\SU(2)$.
The main figure demonstrates that the correlations for the two gauge groups can be scaled on top of each other if, in addition to the mentioned time rescalings, correlations are multiplied by $N_c/(d_{N_c}(N_c^2-1))$.
The variable $d_{N_c}$ counts real degrees of freedom of the Polyakov loop trace and is defined as $d_{N_c}=2$ for $N_c\geq 3$ and $d_2=1$.
This overall rescaling factor will be derived from perturbation theory in \Cref{Sec:PerturbNcDependence}.
The observation that, in the main panel of \Cref{Fig:NcDependence}(b), all of the properly rescaled curves lie on top of each other at sufficiently high momenta is, together with the prescaling results, a strong indication for the applicability of the perturbative analysis of \Cref{Sec:Analytics}.

We emphasize that the perturbative arguments leading to the QCD kinetic theory of \cite{Arnold:2002zm} are only well-defined for sufficiently hard modes above the screening scale $p \gtrsim m_D$ as long as occupancies are not too large: $g^2 f \lesssim 1$. 
These conditions are not satisfied at low momenta, and the occupation number at zero momentum does not need to follow \Cref{eq:Boltzmann}. 
The observed growth of the zero mode in terms of Polyakov loop correlations is, indeed, inconsistent with the scaling ansatz \Cref{eq:prescalingPP} that would predict a decreasing trend instead, and its $N_c$-independent evolution does not seem to agree with predictions from the Boltzmann equation or the analysis of \Cref{Sec:Analytics}. 
These numerical indications confirm that, in contrast to the direct energy cascade, the condensate formation dynamics constitutes an intrinsically nonperturbative phenomenon.

\onecolumngrid

\section{Perturbative expectations}\label{Sec:Analytics}

Given the weak couplings of the simulations, it remains to investigate to what extent the observed self-similar prescaling cascade in Polyakov loop correlations and their $N_c$ scaling can be understood analytically, based on perturbative expectations and known behavior of gauge field correlations.
We first derive a leading-order expression for Polyakov loop correlations in terms of gauge fields, then deduce its perturbative dependence on $N_c$, and finally connect them to the asymptotic scaling exponents $\alpha_\infty$ and $\beta_\infty$.

\subsection{Polyakov loop correlations from gauge field correlations}\label{Sec:AnalyticsPolyakovCorrelationsFromGauge}
We carry out the perturbative derivation in the smooth setting, where the space-time is given by $\rr\times T^3$, $\rr$ denoting a continuous time direction and $T^3$ the spatial 3-torus $T^3 = (\rr\mod L_s)^3$.
This implies spatially periodic boundary conditions as for the lattice setting, which are crucial for the construction of spatial Polyakov loops.
We consider gauge groups $\SU(N_c)$ for general $N_c$ in this analysis, and gauge fields without the map $A\mapsto gA$, which we applied for the simulations.
Then the Polyakov loop trace reads
\begin{equation}
    P(t,\xb) = \frac{1}{N_c}\Tr\, \Pscr\, \exp\bigg\{-ig\oint_0^{L_s} \dd x_2\, A_2(t,\xx)\bigg\}\,.
\end{equation}
Its connected two-point correlations become
\begin{equation}\label{eq:2ptCorrelationsCmooth}
    \langle P(t,\xb)P^\dagger(t,\yb)\rangle_c =\; \frac{1}{N_c^2}\langle \Tr\,\Pscr\,\exp\bigg\{-ig\oint_0^{L_s} \dd x_2\, A_2(t,\xx)\bigg\} \Tr\,\Pscr^{-1}\,\exp\bigg\{ig\oint_0^{L_s} \dd y_2\, A_2(t,\yy)\bigg\}\rangle -\langle P(t,\xb)\rangle \langle P^\dagger(t,\yb)\rangle\,,
\end{equation}
where $\Pscr^{-1}$ indicates path ordering along the negative 2-direction.
We can rewrite the disconnected contributions to \Cref{eq:2ptCorrelationsCmooth} using the cluster decomposition theorem~\cite{van1976stochastic,DiGiacomo:2000irz}:
\begin{align}
    \langle P(t,\xb)\rangle &= \; \frac{1}{N_c}\Tr\, \exp \bigg\{\sum_{n=1}^\infty \frac{(-ig)^n}{n!}\oint_0^{L_s}\dd x_2^{(1)}\cdots \oint_0^{L_s}\dd x_2^{(n)} \langle A_2(t,\xx^{(1)})\cdots A_2(t,\xx^{(n)})\rangle_c\bigg\}\nonumber\\
    &=\;1 - \frac{g^2}{2N_c}\oint_0^{L_s}\dd x_2^{(1)} \oint_0^{L_s}\dd x_2^{(2)} \langle \Tr( A_2(t,\xx^{(1)}) A_2(t,\xx^{(2)}))\rangle_c + \Ocal(g^3)\,,\label{eq:Polyakov1ptClusterDecomp}
\end{align}
where $\xx^{(i)}=(x_1,x_2^{(i)},x_3)$ and the gauge field correlations are connected in the $A_2$.
In \Cref{eq:Polyakov1ptClusterDecomp} we further assumed that $\langle A_2(t,\xx)\rangle = 0$, motivated through Elitzur's theorem~\cite{Elitzur:1975im}, which is consistent with our spatially homogeneous setup.

To further process the first term on the right-hand side of \Cref{eq:2ptCorrelationsCmooth}, we use a result from Lie group representation theory.
Let $\dd \mu$ denote the Haar measure of $\SU(N_c)$, normalized to $\mu(\SU(N_c)) = 1$, and $f:\cc^{N_c}\to \cc^{N_c}$ be a linear map.
Then,%
\footnote{This follows for instance from Lemma 4.42 in~\cite{kirillov2008introduction}, which states that for $G$ a compact finite-dimensional Lie group, $\dd \nu$ its Haar measure, $\rho$ an irreducible finite-dimensional complex representation of $G$ and $f:\rho\to \rho$ a linear map: $\int_G \dd \nu(g) \, \rho(g) f \rho(g)^{-1} = (\Tr \, f / \dim  \rho) \, \mathrm{id}_{\rho}$.
In our case, $G=\SU(N_c)$ and $\rho$ is given by the fundamental representation of $\SU(N_c)$, which is indeed irreducible.
}
\begin{equation}\label{eq:LieTraceIdentity}
    \int_{\SU(N_c)} \dd \mu(V)\, V f V^\dagger = \frac{\Tr \, f}{ N_c}\, 1_{N_c\times N_c}\,,
\end{equation}
where $1_{N_c\times N_c}$ denotes the $N_c\times N_c$ unit matrix.
Using \Cref{eq:LieTraceIdentity}, we obtain 
\begin{align}
    &\frac{1}{N_c^2}\langle \Tr\,\Pscr\,\exp\bigg\{-ig\oint_0^{L_s} \dd x_2\, A_2(t,\xx)\bigg\} \Tr\,\Pscr^{-1}\,\exp\bigg\{ig\oint_0^{L_s} \dd y_2\, A_2(t,\yy)\bigg\}\rangle\nonumber\\
    &=\; \frac{1}{N_c}\int_{\SU(N_c)}\dd \mu(V) \int_{\SU(N_c)}\dd \mu(W)\, \Tr \,\langle V \Pscr_{(t,\xb_0)\to (t,\xb_0)} \exp\bigg\{-ig\oint_0^{L_s}\dd x_2\, A_2(t,\xx)\bigg\} V^\dagger\nonumber\\
    &\qquad\qquad\qquad\qquad\qquad\qquad\qquad\qquad\qquad \times  W \Pscr_{(t,\yb_0)\to (t,\yb_0)}^{-1} \exp\bigg\{ig\oint_0^{L_s}\dd y_2\, A_2(t,\yy)\bigg\} W^\dagger\rangle\,,\label{eq:TraceIdentityRewriting2pt}
\end{align}
where $\xb_0=(x_1,0,x_3)$, $\yb_0=(y_1,0,y_3)$, $\Pscr_{(t,\xb_0)\to (t,\xb_0)}$ indicates path ordering along the positive 2-direction, starting and ending at $(t,\xb_0)$, and $\Pscr_{(t,\yb_0)\to (t,\yb_0)}^{-1}$ indicates path ordering along the negative 2-direction, starting and ending at $(t,\yb_0)$.
The choice of 0 in $\xb_0$ and $\yb_0$ is without loss of generality; independently, two other real numbers in the interval $[0,L_s)$ may equivalently be chosen.

Assuming $\xb\neq \yb$, the $\SU(N_c)$ integrals in \Cref{eq:TraceIdentityRewriting2pt} can be viewed as an average over all gauge transformations of the quantity
\begin{equation}
    \frac{1}{N_c} \Tr \,\langle \Pscr_{(t,\xb_0)\to (t,\xb_0)} \exp\bigg\{-ig\oint_0^{L_s}\dd x_2\, A_2(t,\xx)\bigg\} \Pscr_{(t,\yb_0)\to (t,\yb_0)}^{-1} \exp\bigg\{ig\oint_0^{L_s}\dd y_2\, A_2(t,\yy)\bigg\}\rangle\,.
\end{equation}
We can therefore rewrite \Cref{eq:TraceIdentityRewriting2pt} as
\begin{equation}\label{eq:FunctionalIntegralRewriting}
    \frac{1}{N_c} \int \Dscr\Gcal \, \Tr\,\langle \Pscr'_{t,\xb_0,\yb_0} \exp \bigg\{-ig \oint_0^{L_s} \dd x_2\, A_2(t,\xx) + ig\oint_0^{L_s}\dd y_2\, A_2(t,\yy)\bigg\}\rangle\,,
\end{equation}
where $\Dscr \Gcal$ indicates a functional integral over all $\SU(N_c)$-valued gauge transformations, (formally) normalized to $\int \Dscr \Gcal = 1$.
$\Pscr'_{t,\xb_0,\yb_0}$ denotes path ordering along two disconnected loops: starting at $(t,\yb_0)$, winding around the negative 2-direction to $(t,\yb_0)$, then starting the second loop at $(t,\xb_0)$, winding around the positive 2-direction and ending at $(t,\xb_0)$.
By cyclicity of the trace, \Cref{eq:FunctionalIntegralRewriting} is independent of the ordering among the two disconnected loops.
In writing the two exponentials in \Cref{eq:TraceIdentityRewriting2pt} as a single one in \Cref{eq:FunctionalIntegralRewriting}, we further employed that the equal-time gauge fields commute at different spatial positions.
If we had defined $P(t,\xb)$ using its real part only, then the right-hand side of \Cref{eq:TraceIdentityRewriting2pt} would include taking real parts as well, which would impede this rewriting as a single exponential.
Employing again the cluster decomposition theorem~\cite{van1976stochastic,DiGiacomo:2000irz}, \Cref{eq:FunctionalIntegralRewriting} reads to lowest order in the coupling $g$:
\begin{align}
    &1 - \frac{g^2}{2N_c} \int \Dscr\Gcal\, \bigg[\oint_0^{L_s}\dd x_2^{(1)} \oint_0^{L_s} \dd x_2^{(2)} \langle \Tr(A_2(t,\xx^{(1)}) A_2(t,\xx^{(2)}))\rangle_c - 2\oint_0^{L_s}\dd x_2 \oint_0^{L_s} \dd y_2\, \langle \Tr(A_2(t,\xx) A_2(t,\yy))\rangle_c  \nonumber\\
    &\qquad\qquad\qquad\qquad\qquad\qquad\qquad\qquad\qquad  + \oint_0^{L_s}\dd y_2^{(1)} \oint_0^{L_s} \dd y_2^{(2)} \langle \Tr(A_2(t,\yy^{(1)}) A_2(t,\yy^{(2)}))\rangle_c \bigg] + \Ocal(g^3)\,.\label{eq:Polyakov2ptPerturb}
\end{align}

Since the Polyakov loop trace expectation value \eqref{eq:Polyakov1ptClusterDecomp} is gauge-invariant, taking its gauge transformation average $\int \Dscr \Gcal$ before a perturbative expansion is an identity map.
Therefore, inserting \Cref{eq:Polyakov1ptClusterDecomp,eq:Polyakov2ptPerturb} into \Cref{eq:2ptCorrelationsCmooth}, we finally obtain the leading-order perturbative expression
\begin{equation}\label{eq:Polyakov2ptConnPerturb}
    \langle P(t,\xb)P^\dagger(t,\yb)\rangle_c = \frac{g^2}{N_c}\int \Dscr \Gcal \oint_0^{L_s}\dd x_2 \oint_0^{L_s} \dd y_2 \, \Tr\, \langle A_2(t,\xx) A_2(t,\yy)\rangle_c + \Ocal(g^3)\,.
\end{equation}
We emphasize that, without the gauge averaging $\int \Dscr \Gcal$, the perturbative approximation would not be gauge invariant.

\twocolumngrid

\subsection{Perturbative $N_c$-dependence}\label{Sec:PerturbNcDependence}

From \Cref{eq:Polyakov2ptConnPerturb}, we can deduce the perturbative leading-order $N_c$-scaling of the connected Polyakov loop correlations for nonzero momenta.
Apart from the explicit prefactor of $1/N_c$, we have 
\begin{align}
\Tr\, \langle A_2(t,\xx) A_2(t,\yy)\rangle_c &=\; \frac{1}{2}\sum_{a=1}^{N_c^2-1} \langle A_2^a(t,\xx)A_2^a(t,\yy)\rangle_c\nonumber\\
&\sim\; N_c^2-1\,,
\end{align}
where we used that the color-specific gauge field correlations $\langle A_2^a(t,\xx)A_2^a(t,\yy)\rangle_c$ contribute equally and are insensitive to $N_c$.
Moreover, while traces over $\SU(2)$-elements are always real-valued, for $N_c\geq 3$ traces over $\SU(N_c)$-elements can also have an imaginary part.
Since the perturbative derivation leading to \Cref{eq:Polyakov2ptConnPerturb} is based on $P(t,\xb)$, which is not restricted to the real part, the gauge average $\int \Dscr \Gcal$ results in a combinatorial prefactor of $d_{N_c}=2$ for $N_c\geq 3$ and $d_2=1$.
Putting these prefactors together, we obtain
\begin{equation}
    \langle P(t,\xb)P^\dagger(t,\yb)\rangle_c\sim d_{N_c}\frac{N_c^2-1}{N_c}\,.
\end{equation}
This is indeed consistent with the findings of \Cref{Fig:NcDependence}(b).

\subsection{Self-similar scaling}\label{Sec:PerturbativeSelfSimilarScaling}
For the chosen initial conditions with large gluonic over-occupation in the infrared, gauge field distributions and correlations approach self-similar (pre-)scaling dynamics in time, which constitutes the direct energy cascade.
In spatially homogeneous scenarios such as ours, we can write for the gluon correlations 
\begin{equation}
\Tr\, \langle A_2 A_2\rangle_c(t,\Delta\xx) = \Tr\, \langle A_2(t,\xx) A_2(t,\yy)\rangle_c\,,
\end{equation}
where $\Delta \xx = \xx-\yy$.
Focusing on self-similar dynamics at asymptotically late times, these can then exhibit the following scaling behavior in the vicinity of the nonthermal fixed point:%
\footnote{This specific form of self-similar scaling follows from the prescaling ansatz \eqref{Eq:PrescalingAnsatzAA} for the gluon correlations in momentum space, a Fourier transform and the behavior of $a(t)$ and $b(t)$ at asymptotically late times.}
\begin{equation}\label{eq:AAcorrelationsScaling}
    \Tr\, \langle A_2 A_2\rangle_c(t,\Delta \xx) = t^{\alpha_\infty-2\beta_\infty} \Tr\, \langle A_2  A_2\rangle_{c,S}(t^{-\beta_\infty}\Delta\xx)\,.
\end{equation}
Here, $\Tr\, \langle A_2  A_2\rangle_{c,S}(\Delta \xx)$ indicates the fixed-point shape of $\Tr\, \langle A_2 A_2\rangle_c(t,\Delta \xx)$.
A reference time scale for the temporal power law has been absorbed into the former.

Insertion of \Cref{eq:AAcorrelationsScaling} into \Cref{eq:Polyakov2ptConnPerturb} yields the self-similar scaling behavior
\begin{equation}\label{eq:Conn2ptCorrelationsPolyakovScaling}
    \langle PP^\dagger\rangle_c(t,\Delta\xb) =t^{\alpha_\infty} \langle P P^\dagger\rangle_{c,S}(t^{-\beta_\infty}\Delta\xb)\,,
\end{equation}
where periodicity of the loop integrals in \Cref{eq:Polyakov2ptConnPerturb} has been employed and $\langle P P^\dagger\rangle_{c,S}$ denotes again the fixed-point shape.
\Cref{eq:Conn2ptCorrelationsPolyakovScaling} can now be conveniently Fourier transformed to momentum space, yielding
\begin{align}
    \langle PP^\dagger\rangle_c(t,\pb) =&\; \int_{\Delta\xb} \langle PP^\dagger\rangle_c(t,\Delta\xb) e^{-i\pb \Delta \xb} \nonumber\\
    =&\; t^{\alpha_\infty+2\beta_\infty}\langle PP^\dagger\rangle_{c,S}(t^{\beta_\infty}\pb)\,.\label{eq:PPScalingPredictionPerturbative}
\end{align}
For the energy cascade in pure Yang-Mills theory out of equilibrium, one has $\alpha_\infty = 4\beta_\infty$ and $\beta_\infty=-1/7$~\cite{Kurkela:2012hp, Schlichting:2012es, Berges:2013fga, AbraaoYork:2014hbk, Boguslavski:2018beu}.
The scaling behavior of \Cref{eq:PPScalingPredictionPerturbative} is consistent with our results discussed in \Cref{Sec:PrescalingUV} for asymptotically late times. 
The prescaling ansatz generalizes these results by substituting $t^{\alpha_\infty} \mapsto a(t)$ and $t^{\beta_\infty} \mapsto b(t)$.

In the derivation, we have implicitly assumed that the scaling behavior \eqref{eq:AAcorrelationsScaling} for the gauge field correlations holds in arbitrary gauge.
While no studies of the nonequilibrium gluon plasma are known to us that employ gauges other than the temporal-axial and spatial Coulomb-type gauges for the considered system, this assumption is \emph{a posteriori} justified by the accurate (pre-)scaling description of the gauge-invariant Polyakov loop correlations and their perturbative relation to the gauge field correlations.

\section{Conclusions}\label{Sec:Conclusions}

In this work, we considered real-time $\SU(N_c)$ lattice gauge theory in the classical-statistical regime governed by weak couplings and large gluon occupations.
We focused on the connected 2-point correlations of spatial Polyakov loop traces, which are gauge-invariant observables related to gluon occupation numbers but distinct from gauge-dependent gluon distribution functions.
Our results confirm that the zero mode of the Polyakov loop correlations exhibits the dynamic formation of a gauge-invariant condensate. 
For nonzero momenta, we observed that the correlations are governed by a self-similar direct cascade consistent with energy conservation. We demonstrated their dynamics to be accurately described by prescaling, where scaling exponents exhibit nontrivial time-dependencies and only asymptotically converge to their universal fixed values that agree with the universal exponents from gauge field correlations.
This leads to a relatively simple composition of Polyakov loop correlation dynamics: the zero mode condenses while nonzero modes transport energy toward the ultraviolet in a direct prescaling cascade, with the condensate fraction growing to dominate the total Polyakov loop occupations over time.

To better understand the prescaling and, in particular, the asymptotic values of the scaling functions, we carried out a detailed perturbative analysis of Polyakov loop correlations.
Utilizing basic results from Lie group representation theory, the perturbative expressions featured averages over gauge transformations to maintain gauge invariance.
We derived the leading-order contributions of gauge field correlations to the Polyakov loop correlations.
Based on the known scaling behavior of the former, this allowed us to consistently describe the asymptotic self-similar behavior of the Polyakov loop correlations and motivated a working ansatz for their prescaling dynamics.

We also examined the $N_c$-dependence of Polyakov loop correlations by comparing results for gauge groups $\SU(2)$ and $\SU(3)$. In the temporal regime studied, dynamics progressed by a factor $(3/2)^2$ faster for $\SU(3)$ than for $\SU(2)$, as predicted by kinetic theory in the classical approximation~\cite{AbraaoYork:2014hbk}.
The nonzero momentum modes scaled with a nontrivial $N_c$-dependent factor, consistent with our perturbative analysis.

Other than for scalar theories, where condensation, as captured by the build-up of zero mode correlations, is often accompanied by an inverse cascade transporting particle number toward the infrared~\cite{Berges:2012us, PineiroOrioli:2015cpb}, we found no evidence for the existence of a similar inverse cascade in the investigated non-Abelian gauge theories for the chosen lattice discretizations.
However, our numerical study does not rule out the possibility of an inverse cascade below the electric screening scale, warranting further studies to clarify whether an inverse cascade exists and, more generally, what the dynamical condensation mechanism is.

We anticipate that spatial Wilson loop correlations would exhibit dynamics similar to those for Polyakov loop correlations due to existing similarities between spatial Polyakov and Wilson loops in far-from-equilibrium conditions, see e.g.~\cite{Berges:2017igc, Berges:2019oun, Berges:2023sbs}.
Furthermore, gauge-invariant condensation, as captured by Wilson loops, bears similarities to scalar $\mathrm{O}(N)$ models near a nonthermal fixed point~\cite{Berges:2019oun}.
Investigating Polyakov loop correlations in anisotropic scenarios may also yield insights relevant to the successful description of collisions of atomic nuclei~\cite{Bjorken:1982qr, Berges:2020fwq}.
For instance, for relativistic scalars, the dynamic condensate formation can be similar for both isotropic and expanding anisotropic systems~\cite{Berges:2015ixa}.
Finally, our results may provide crucial indications for novel effective or kinetic gauge-invariant descriptions of gluon dynamics based on spatial Polyakov loop correlations. 

In conclusion, by employing gauge-invariant observables, our work provides an important step toward a better understanding of the nonequilibrium dynamics in gluonic systems and their role in the thermalization of the quark-gluon plasma.
Despite similarities with scalar theories, the simple picture of condensation and ultraviolet cascade observed here for $\SU(N_c)$ gauge theories using Polyakov loop correlations may distinguish the gluon dynamics from relativistic scalar field theories, which typically exhibit a pronounced inverse cascade driving condensate formation. 
Further studies are needed to confirm or refute this intriguing picture.

\begin{acknowledgments}

We thank J.~Berges, T.~Butler, L.~de~Bruin, J.M.~Pawlowski and P.~Radpay for valuable discussions and collaboration on related studies. 
This work is part of and funded by the Deutsche Forschungsgemeinschaft (DFG, German Research Foundation) under Germany’s Excellence Strategy EXC 2181/1–390900948 (the Heidelberg STRUCTURES Excellence Cluster) and the Collaborative Research Centre, Project-ID No. 273811115, SFB 1225 ISOQUANT.
This research was funded in part by the Austrian Science Fund (FWF) [10.55776/P34455].
We acknowledge support by the Interdisciplinary Center for Scientific Computing (IWR) at Heidelberg University, where part of the numerical work has been carried out.
\end{acknowledgments}

\appendix

\begin{figure}[th!]
    \centering
	\includegraphics[width=0.95\columnwidth]{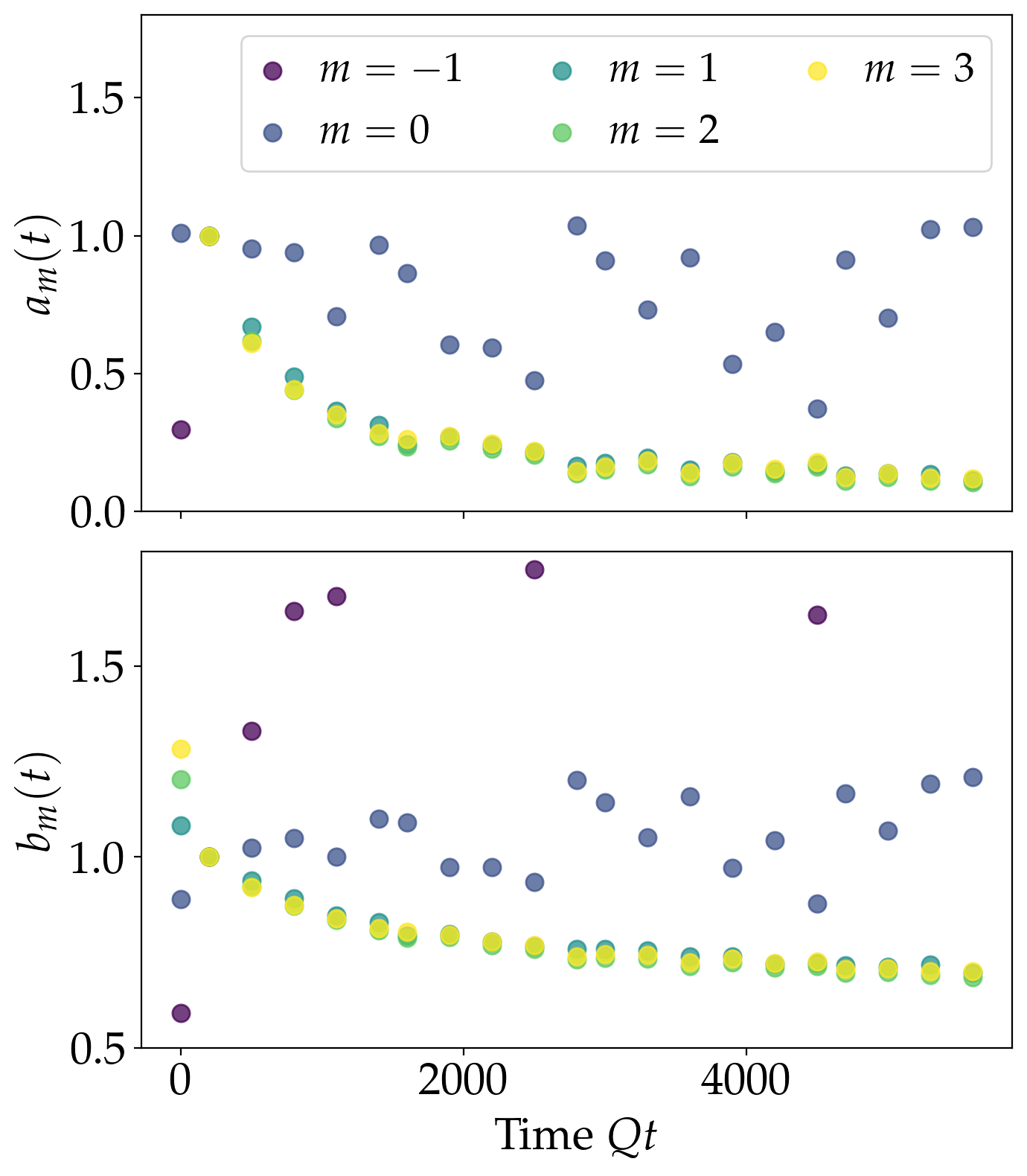}
	\caption{Prescaling factors $a_m(t),b_m(t)$ extracted from different moments $n_m(t)$ of the Polyakov correlations. 
 The higher moments $m\geq 1$ probe the momentum range of the direct cascade and are observed to converge quickly in time, reflecting prescaling.}\label{Fig:AppPrescaling}
\end{figure}

\begin{figure}[th!]
    \centering
	\includegraphics[width=0.95\columnwidth]{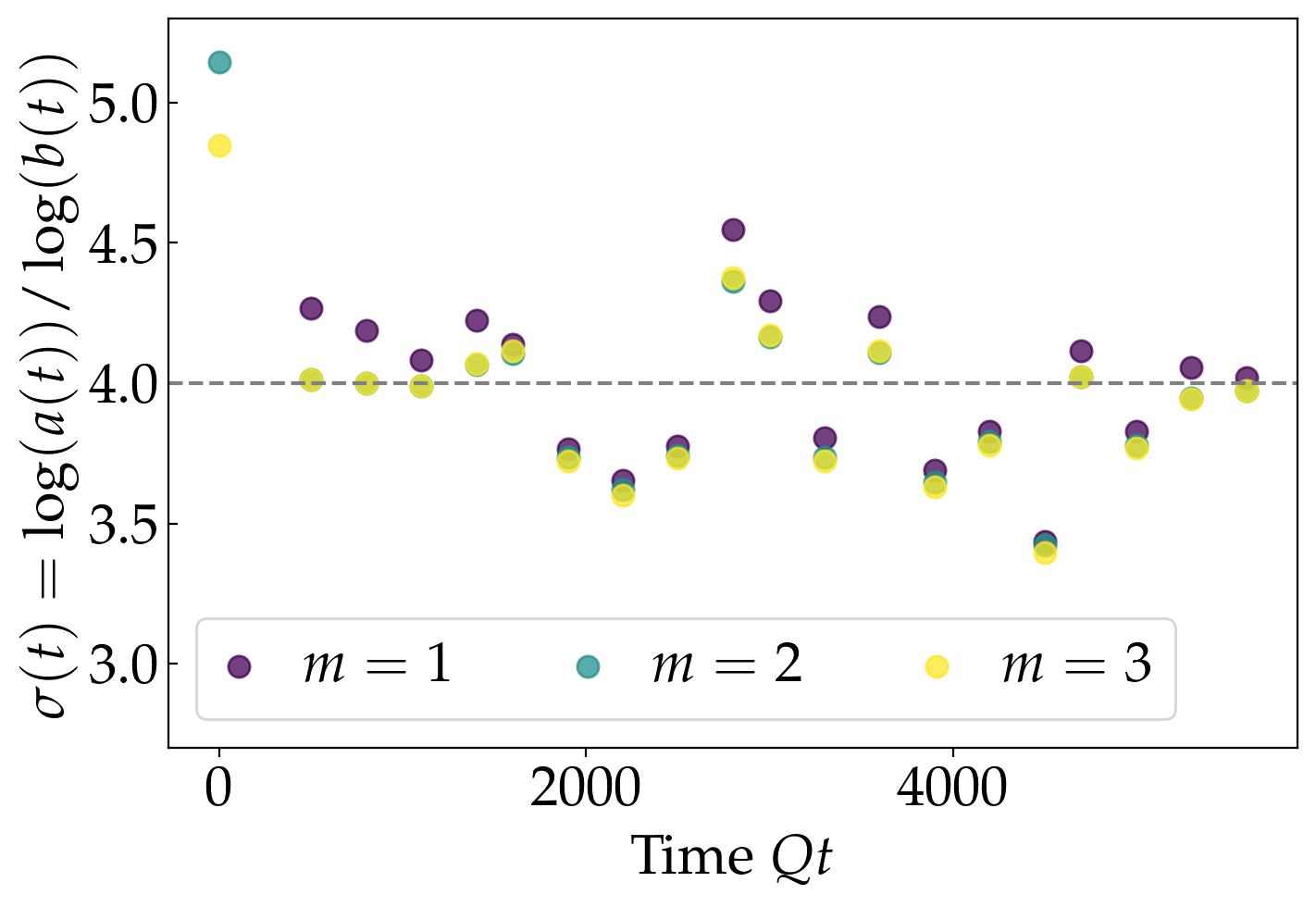}
	\caption{Logarithmic ratio of the prescaling factors $a(t),b(t)$ extracted from different moments $n_m(t)$ of the Polyakov correlations. 
 The logarithmic ratio $\sigma(t)=4$ (horizontal dashed line) reflects mode energy conservation of relativistic gluons in three spatial dimensions.
 }\label{Fig:AppEnergyConservation}
\end{figure}

\section{Prescaling functions and their extraction}
\label{appendix:Prescaling}
The prescaling functions $a(t),b(t)$ are extracted according to the protocol presented in~\cite{Heller:2023mah}. Starting from \Cref{eq:prescalingPP}, we introduce moments of the Polyakov loop correlations
\begin{align}
    n_m(t) &= \int \frac{d^2 \bar{\mathbf{p}}}{(2\pi)^2} p^m \langle PP^\dagger \rangle_c(t,p) \\
    &= a(t) b^{-m}(t) n_m(t_0)\,.
\end{align}
We then extract the prescaling factors $a(t),b(t)$ from combinations of the moments according to
\begin{subequations}
\label{eq.nonexp_AB_moments}
\begin{align}
a_m(t) &=\left[\frac{n_m^{1/m}(t)}{n_{m+1}^{1/(m+1)}(t)} \frac{n^{1/(m+1)}_{m+1}(t_0)}{n^{1/m}_m(t_0)} \right]^{m(m+1)}\,,\\
b_m(t) &= \frac{n_m(t)}{n_{m+1}(t)} \frac{n_{m+1}(t_0)}{n_m(t_0)}\,,
\end{align}
\end{subequations}
with $a_m(t_0)=b_m(t_0)=1$, which are shown in \Cref{Fig:AppPrescaling}. 
While lower moments mix infrared and ultraviolet dynamics, prescaling factors extracted from $m\geq 1$ converge quickly in time, reflecting the emergence of prescaling for the direct cascade in the ultraviolet. 
For the prescaling figures with $a(t),b(t)$ in the main text, we average the prescaling factors obtained from $m=1,2,3$.

 The direct cascade encodes energy transport of relativistic gluons in three spatial dimensions, which is reflected in $a(t)=b(t)^4$ as is visible in \Cref{Fig:AppEnergyConservation}.

\nocite{spitz_2025_15205964}

\bibliography{literature}

\end{document}